\documentclass[reprint,prl,superscriptaddress]{revtex4-1}
\usepackage{graphicx}
\usepackage{epsfig}
\usepackage{color}
\usepackage{bm}
\usepackage{amsmath}

\begin{document}

\title{Possibility of an unconventional spin state of Ir$^{4+}$ in~$\mathbf{Ba_{21}Ir_9O_{43}}$~single~crystal}

\author{L.~Yang}
\affiliation{Laboratory of Physics of Complex Matter, Institute of Physics, Ecole Polytechnique F\'{e}derale de Lausanne (EPFL), CH-1015 Lausanne, Switzerland}
\affiliation{Laboratory for Quantum Magnetism, Institute of Physics, Ecole Polytechnique F\'{e}derale de Lausanne (EPFL), CH-1015 Lausanne, Switzerland}
\author{M.~Jeong}
\affiliation{Laboratory for Quantum Magnetism, Institute of Physics, Ecole Polytechnique F\'{e}derale de Lausanne (EPFL), CH-1015 Lausanne, Switzerland}
\author{A.~Arakcheeva}
\affiliation{Laboratory of Physics of Complex Matter, Institute of Physics, Ecole Polytechnique F\'{e}derale de Lausanne (EPFL), CH-1015 Lausanne, Switzerland}
\author{I.~\v Zivkovi\'c}
\affiliation{Laboratory for Quantum Magnetism, Institute of Physics, Ecole Polytechnique F\'{e}derale de Lausanne (EPFL), CH-1015 Lausanne, Switzerland}
\author{B.~N\'{a}fr\'{a}di}
\affiliation{Laboratory of Physics of Complex Matter, Institute of Physics, Ecole Polytechnique F\'{e}derale de Lausanne (EPFL), CH-1015 Lausanne, Switzerland}
\author{A.~Magrez}
\affiliation{Crystal Growth Facility, Institute of Physics, Ecole Polytechnique F\'{e}derale de Lausanne (EPFL), CH-1015 Lausanne, Switzerland}
\author{A.~Pisoni}
\affiliation{Laboratory of Physics of Complex Matter, Institute of Physics, Ecole Polytechnique F\'{e}derale de Lausanne (EPFL), CH-1015 Lausanne, Switzerland}
\author{J.~Jacimovic}
\affiliation{Laboratory of Physics of Complex Matter, Institute of Physics, Ecole Polytechnique F\'{e}derale de Lausanne (EPFL), CH-1015 Lausanne, Switzerland}
\author{V.~M.~Katukuri}
\affiliation{Institute of Physics, Ecole Polytechnique F\'{e}derale de Lausanne (EPFL), CH-1015 Lausanne, Switzerland}
\author{S. Katrych}
\affiliation{Laboratory of Physics of Complex Matter, Institute of Physics, Ecole Polytechnique F\'{e}derale de Lausanne (EPFL), CH-1015 Lausanne, Switzerland}
\author{N.~E.~Shaik}
\affiliation{Laboratory for Quantum Magnetism, Institute of Physics, Ecole Polytechnique F\'{e}derale de Lausanne (EPFL), CH-1015 Lausanne, Switzerland}
\author{O.~V.~Yazyev}
\affiliation{Institute of Physics, Ecole Polytechnique F\'{e}derale de Lausanne (EPFL), CH-1015 Lausanne, Switzerland}
\author{L.~Forr\'{o}}
\affiliation{Laboratory of Physics of Complex Matter, Institute of Physics, Ecole Polytechnique F\'{e}derale de Lausanne (EPFL), CH-1015 Lausanne, Switzerland}
\author{H.~M.~R\o nnow}
\email{henrik.ronnow@epfl.ch}
\affiliation{Laboratory for Quantum Magnetism, Institute of Physics, Ecole Polytechnique F\'{e}derale de Lausanne (EPFL), CH-1015 Lausanne, Switzerland}

\begin{abstract}
We report the synthesis of single crystals of a novel layered iridate $\mathrm{Ba_{21}Ir_9O_{43}}$, and present the crystallographic, transport and magnetic properties of this material. The compound has a hexagonal structure with two iridium oxide layers stacked along the $c$ direction. One layer consists of a triangular arrangement of $\mathrm{Ir_2O_9}$ dimers while the other layer comprises two regular octahedra and one triangular pyramid, forming inter-penetrated triangular lattices. The resistivity as a function of temperature exhibits an insulating behavior, with a peculiar $T^{-3}$ behavior. Magnetic susceptibility shows antiferromagnetic Curie-Weiss behavior with $\Theta_\mathrm{CW} \simeq -90$~K while a magnetic transition occurs at substantially lower temperature of 9 K. We discuss possible valence states and effective magnetic moments on Ir ions in different local environments, and argue that the Ir ions in a unique triangular-pyramidal configuration likely carry unusually large magnetic moments.
\end{abstract}

\date{\today}

\maketitle

There is considerable interest in studying correlated $5d$-electron transition metal oxides with strong spin-orbit coupling (SOC)~\cite{Pesin10NaturePhys, Witczak13ARCMP, Rau16ARCMP}. Recent focus has been on new iridate compounds where experimental discoveries have strongly challenged theoretical predictions. For example, in pentavalent (Ir$^{5+}$) $5d^4$ iridates a strong SOC is predicted to realize a nonmagnetic $J=0$ state \cite{Khomskii}, but substantial magnetic moments have been found in $\mathrm{Ba_2YIrO_6}$ \cite{Dey16PRB} and especially in $\mathrm{Sr_2YIrO_6}$ \cite{Cao14PRL}, where structural distortion of the octahedral environment leads to 0.91~$\mu_\mathrm{B}$/Ir. Furthermore, a possible realization of a spin-orbital liquid with Ir$^{5+}$ has been discussed for $6H$ hexagonal $\mathrm{Ba_3ZnIr_2O_9}$ powder \cite{Nag15arxiv}.

In tetravalent (Ir$^{4+}$) 5$d^5$ iridates, such as $\mathrm{Sr_2IrO_4}$~\cite{Cao98PRB, Kim08PRL, Kim09Science}, a complex spin-orbit entangled $J_\mathrm{eff}=1/2$ state is realized, with a reduced bandwidth due to the splitting from the $J_\mathrm{eff}=3/2$ states, which leads to the localization of electrons even by a relatively small on-site Coulomb interaction compared to their $3d$ counterparts. Additional effects of SOC include XY anisotropy \cite{Vale15PRB, Katukuri14PRX}, locking of the magnetic moment to the correlated rotation of oxygen octahedra in $\mathrm{Sr_2IrO_4}$~\cite{Boseggia13JPCM}, anisotropic dimer excitations in $\mathrm{Sr_3Ir_2O_7}$~\cite{Sala15PRB}, and Kitaev type of magnetic interactions in $\mathrm{Na_2IrO_3}$~\cite{Gegenwart15NaturePhys}. Structural distortions from the ideal octahedral environment have been observed in the pyrochlore family of the type $R_2\mathrm{Ir_2O_7}$ ($R$=rare-earth) where trigonal compression induces a reconstruction of energy levels~\cite{Uematsu15PRB,Hozoi14PRB}.

So far, to the best of our knowledge, there have been very few reports~\cite{Kanungo15AngewandteChemie} of oxides containing Ir$^{4+}$ or Ir$^{5+}$ ions in an environment different from octahedral. Following the observations of profound influence of structural effects onto the energy scheme of iridium ions and consequently the resulting spin-orbital state, it can be expected that non-octahedral ligand coordination could result in different properties.

\begin{figure*}
	\centering
	\includegraphics[width=1\textwidth]{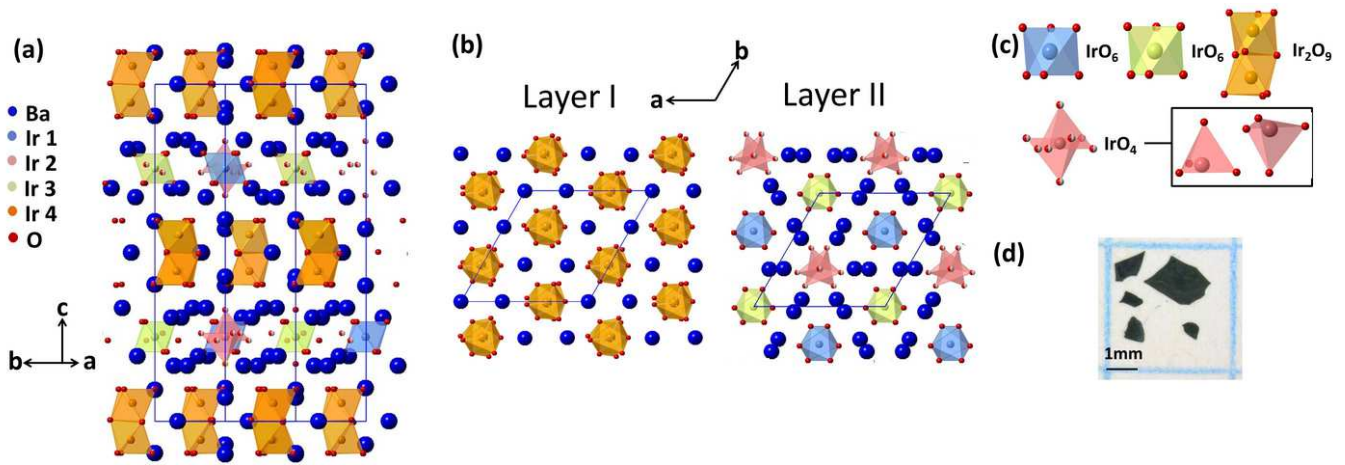}
	\caption{(a) Crystal structure of $\mathrm{Ba_{21}Ir_9O_{43}}$ where a unit cell is drawn by solid line. (b) Two different layers of triangular lattice Ir-O polyhedra, labeled as Layer I $\mathrm{[Ba_9Ir_6O_{27}]^{-6}}$ (left) and Layer II $\mathrm{[Ba_{12}Ir_3O_{16}]^{+6}}$ (right). (c) The Ir-O polyhedra: blue and green $\mathrm{IrO_6}$ octahedra are geometrically identical within one standard deviation; face-sharing orange octahedra form a $\mathrm{Ir_2O_9}$ dimer; red polyhedron is shown as a statistical superposition of two possible $\mathrm{IrO_4}$ triangular pyramid, having the same geometry but opposite in orientation along $c$ direction, with {50 \%} occupancy on every oxygen sites. (d) Representative single crystals of $\mathrm{Ba_{21}Ir_9O_{43}}$.}\label{Fig1}
\end{figure*}

Here we report the synthesis, crystal structure, transport, and magnetic properties of single crystals of a novel layered iridate $\mathrm{Ba_{21}Ir_9O_{43}}$. It consists of layers of a triangular lattice of $\mathrm{Ir_2O_9}$ dimers, much alike in $\mathrm{Ba_3ZnIr_2O_9}$, that are intercalated by another layer of a triangular lattice containing three different iridium environments: two regular octahedra and a triangular pyramid (see Fig.~\ref{Fig1}(a-c) for the structure). We conclude that the triangular pyramid represents a novel environment for an Ir$^{4+}$ ion with a high-spin state $J_\mathrm{eff}=5/2$.

Single crystals of $\mathrm{Ba_{21}Ir_9O_{43}}$ were grown in two steps. Firstly, a mixture of powders with the molar ratio of $\mathrm{BaCO_3}$ : $\mathrm{IrO_2}$ = 13 : 2 was heated for 2 days at 1000~$^{\circ}$C and cooled down slowly (2~$^{\circ}$C/h) to 700~$^{\circ}$C in $\mathrm{Al_2O_3}$ crucible. After the reaction, black powder was obtained and pressed into a pellet. Secondly, the mixture of the weight ratio of the pellet : $\mathrm{K_2CO_3}$ powder flux = 1 : 20 was heated for 2 days at 1150~$^{\circ}$C and cooled down to 800~$^{\circ}$C at 2~$^{\circ}$C/h in $\mathrm{Al_2O_3}$ crucible. Black-colored single crystals of $\mathrm{Ba_{21}Ir_9O_{43}}$ were obtained with plate-like shape after the reaction, as shown in Fig.~\ref{Fig1}(d), where the maximum dimensions were $2\times 2\times 0.1$ mm$^3$. The direction perpendicular to the plate coincides with the crystallographic $c$ axis of the hexagonal system. The chemical composition was determined using an Energy Dispersive X-Ray detector (EDX, Oxford Instruments EDX X-MAX), which yielded the molar ratios Ba : Ir of 69.5(4) : 30.6(7).

Single-crystal x-ray diffraction data was collected at room temperature using a SuperNova diffractometer equipped with a CCD detector and high intensity micro-focus x-ray source (Mo $K\alpha$ radiation). Data reduction and analytical absorption correction were done using CrysAlispro software \cite{Dui03JAC}. The crystal structure was solved by Superflip algorithm \cite{Oszlanyi04AC} and refined using JANA2006~\cite{Petricek14ZK}. Resistivity was measured on a single plate of crystal using a standard four-probe configuration with DC current. Bulk magnetic measurements were performed using a Superconducting Quantum Interference Device (SQUID) magnetometer. A dozen of plate-like crystals with total mass of 9.4~mg were stacked with common $c$ axis but random orientation in their $ab$ plane. Electron spin resonance (ESR) measurements were performed on the stacked crystals using a Bruker Elexys E500 continuous wave spectrometer at 9.4~GHz, equipped with a He-gas flow cryostat.

Figure~\ref{Fig1}(a) shows the crystal structure which belongs to the hexagonal system. The details of the data collection and structure refinement are listed in Table~\ref{table1}, and further information is available in Supplemental Material \cite{SM}. The structure consists of two different layered blocks. For the layer I, the central structural feature is the formation of $\mathrm{Ir_2O_9}$ dimers composed of two face-sharing $\mathrm{IrO_6}$ octahedra (Ir4), as shown in Fig.~\ref{Fig1}(b) and (c). Such dimers are typical for 6$H$-perovskite-type oxides \cite{Doi04JPCM, Burbank48AC}. Layer II consists of isolated $\mathrm{IrO_6}$ octahedra (Ir1 and Ir3) and $\mathrm{IrO_4}$ triangular pyramid (Ir2). The $\mathrm{IrO_4}$ pyramids appear in two possible orientations, that is, an apical oxygen pointing either upward or downward, in statistically equal distribution. The chemical formula of $\mathrm{Ba_{21}Ir_9O_{43}}$ was obtained from the fully occupied positions for all sites of all Ir atoms with the accuracy 1.00(1). The molar ratio of Ba/Ir from the formula 21/9=2.33 is close to the value 69.5(4)/30.6(7) = 2.27(7) from the EDX analysis.

\begin{table}[b]      
	\caption{Details of the data collection and structure refinement for $\mathrm{Ba_{21}Ir_9O_{43}}$ obtained at room temperature using Mo $K\alpha$ radiation with $\lambda$ = 0.71073~{\AA}.}
	\begin{tabular}{c c}
		\hline\hline
		Empirical formula &  $\mathrm{Ba_{21}Ir_9O_{43}}$\\
		\hline
		Formula weight (g/mol) &  5302.07\\
		space group &  $\mathrm{P6_322}$\\
		Unit cell parameters ({\AA}) &  $a=b=10.5102(9)$\\		
		&  $c=25.1559(19)$\\
		Volume ({$\mathrm{\AA^3}$}) & 2406.5\\
		Z & 2\\
		Calculated density ({$\mathrm{g/cm^3}$}) & 7.317\\
		Reflections collected/unique & 6486/1188\\
		Data/restraints/parameters & 1188/0/81\\
		Final $R$ indices [$I$ $>$ $3\sigma(I)$] & $R$ = 0.0629, $wR$ = 0.0586\\
		Goodness-of-fit & 1.46\\
		\hline\hline
	\end{tabular}\label{table1}
\end{table}

\begin{figure}
	\centering
	\includegraphics[width=0.5\textwidth]{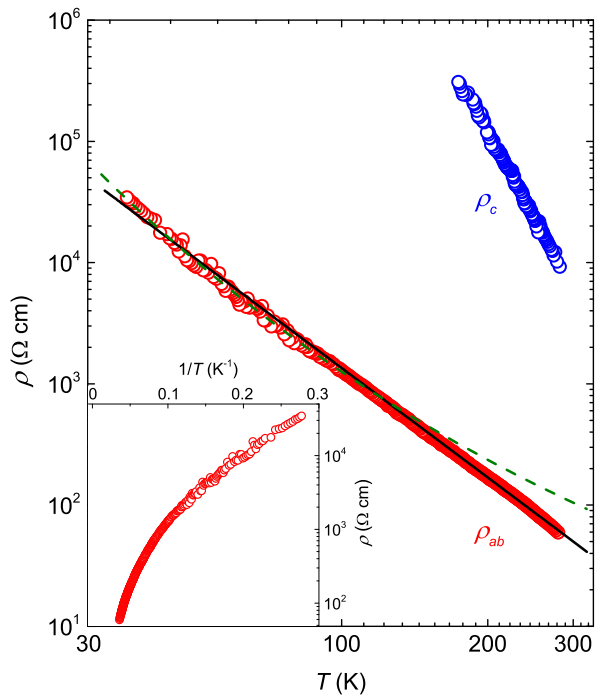}
	\caption{Log-log plot of in-plane ($\rho_{ab}$) and out-of-plane ($\rho_{c}$) resistivity as a function of temperature where dashed and solid lines represent the two-dimensional VRH model and $\rho\propto T^{-3}$ behavior, respectively. Inset shows an Arrhenius plot of $\rho_{ab}$.}\label{Fig2}
\end{figure}

The electrical resistivity $\rho$ as a function of temperature along the in-plane ($\rho_{ab}$) and the out-of-plane ($\rho_c$) directions are shown in Fig.~\ref{Fig2}. At room temperature the magnitudes are $\rho_{ab}=60~\Omega$cm and $\rho_{c}=10^4~\Omega$cm, respectively. Both $\rho_{ab}$ and $\rho_{c}$ increase monotonically as temperature decreases, revealing the insulating nature of this compound. $\rho_c$ was measured only above 150~K due to high resistance.

The $\rho_{ab}$ could not be fitted by a thermally activated exponential behavior, $\rho\propto \exp(E_a/k_\mathrm{B}T)$, where $E_a$ is an activation energy. This is evident from an Arrhenius plot shown in the inset of Fig.~\ref{Fig2} where no linear regime is found. On the other hand, we note that the resistivity in a number of iridates such as $\mathrm{Sr_3Ir_2O_7}$ \cite{Cao02PRB}, $\mathrm{NaIrO_3}$ \cite{Bremholma11JSSC, Jenderka13PRB}, $\mathrm{Ba_2IrO_4}$ \cite{Okabe13PRB}, and $\mathrm{Sr_2IrO_4}$ \cite{Lu14APL} has been found to fit to a variable range hopping (VRH) model. Indeed, the low-temperature resistivity below 120~K can be fitted to $\rho(T)=\rho_0\mathrm{exp}(T_0/T)^{1/3}$, which is known as the two-dimensional VRH model, and the result is shown by dashed line in Fig.~\ref{Fig2}. However, we also discover that $\rho_{ab}$ follows remarkably well a power law as $\rho_{ab}\propto T^{-3}$ as shown by solid line in a log-log plot of Fig.~\ref{Fig2}. While this might be a coincidence, it could suggest that $\rho$ is controlled by unconventional scaling. Results of Seebeck coefficient measurements are given in Supplemental Materials \cite{SM}.

\begin{figure}
	\centering
	\includegraphics[width=0.5\textwidth]{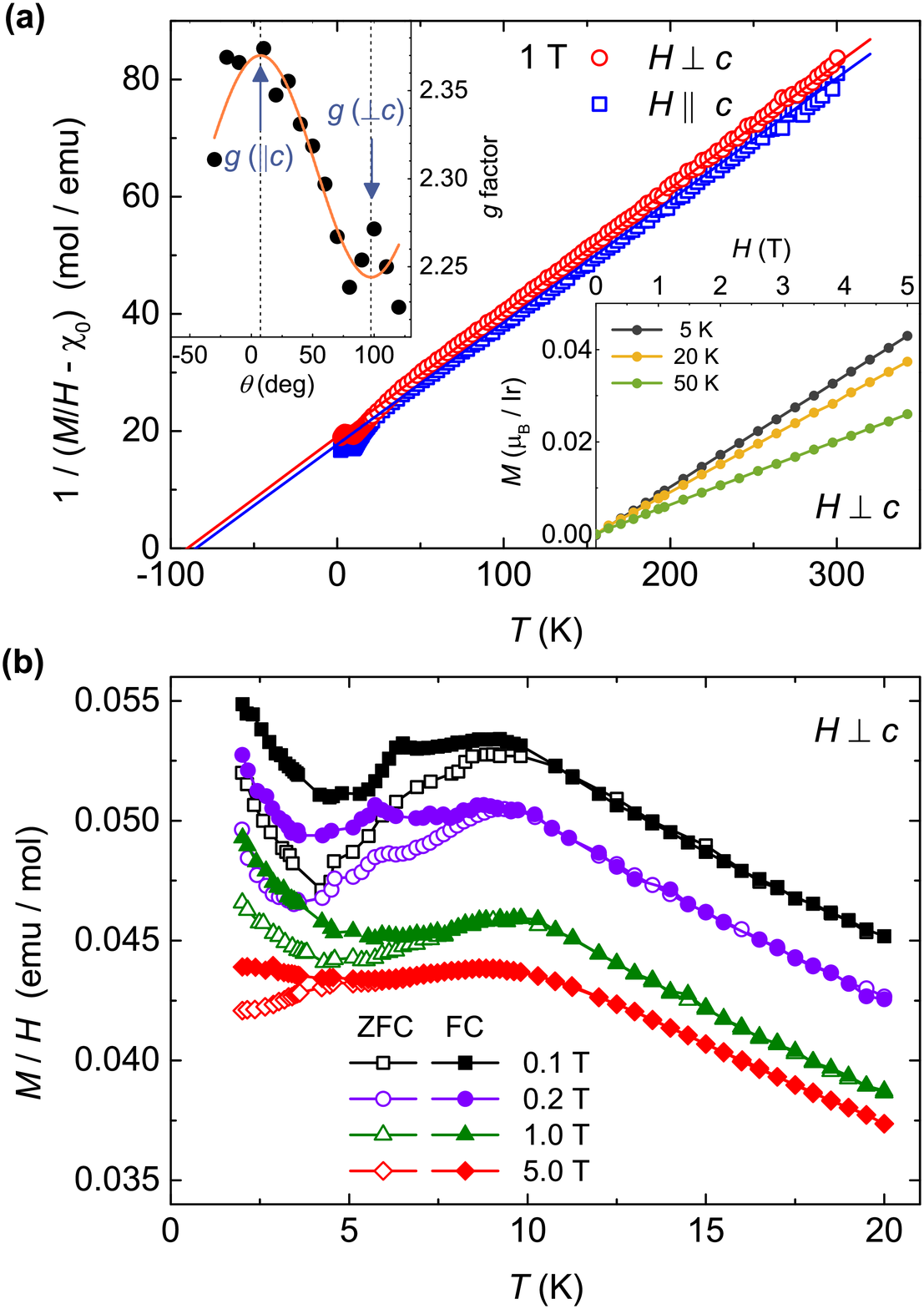}
	\caption{(a) Inverse magnetic susceptibility $ 1/(M/H-\chi_0) $ versus temperature in $H$ = 1~T applied perpendicular and parallel to the $c$ axis, respectively. Lower-right inset: isothermal magnetization as a function of field for different temperatures. Upper-left inset: angular dependence of the $g$ factor at room temperature. (b) The susceptibility as a function of temperature in various applied fields.}\label{Fig3}
\end{figure}

Figure~\ref{Fig3}(a) shows the inverse magnetic susceptibility $ 1/(M/H-\chi_0) $, where $M$ is magnetization, as a function of temperature for an applied field $H$ = 1~T parallel (square) and perpendicular (circles) to the $c$ axis. The high temperature part follows Curie-Weiss behavior, as shown by the solid lines. Overall, the magnetic behavior appears qualitatively similar for both $H\parallel c$ and $H \perp c$. The lower-right inset of Fig.~\ref{Fig3}(a) shows the magnetization as a function of field for $H\perp c$ at different temperatures. The $M(H)$ curve becomes linear with $H$ above 1.5 T in the measured temperature range.

The data above 100~K was fitted by a Curie-Weiss law, $\chi(T)=C/(T-\Theta_\mathrm{CW})+\chi_0$, yielding for $H \perp c$ a Curie constant $C = 4.72(9)$ emu/mol$\,$K, corresponding to an effective moment $\mu_\mathrm{eff} = 2.05(2) \mu_\mathrm{B}$/Ir, a Curie-Weiss temperature $\Theta_\mathrm{CW} = -90(2)$~K and a temperature independent term $\chi_0 = -6.4 \times 10^{-3}$ emu/mol. For $H\parallel c$ the fit gave $C$ = 4.8(1) emu/mol K, $\mu_\mathrm{eff} = 2.07(3) \mu_\mathrm{B}$/Ir, $\Theta_\mathrm{CW} = - 85(3)$~K and $\chi_0 = 1.2(3) \times 10^{-3} $ emu/mol. From the negative value of $\Theta_\mathrm{CW}$ we infer that the dominant magnetic interaction is antiferromagnetic. The temperature independent term $\chi_0$ is rather large and anisotropic, but similar large and anisotropic values have been reported for other iridates, such as $\mathrm{NaIrO_3}$ \cite{Bremholma11JSSC}, $\mathrm{Pr_2Ir_2O_7}$ \cite{Chou04PRB}, and $\mathrm{Ba_8Al_2IrO_{14}}$ \cite{Yang15InorgChem}. We encourage a theoretical scrutiny of these unusual $\chi_0$ effects.

The low temperature behavior of $\chi(T)$ in different fields is shown in Fig.~\ref{Fig3}(b). Most notably, a splitting for zero-field-cooled (ZFC) and field-cooled (FC) is observed below the 9~K maximum while the bifurcation point shifts to lower temperature as the field strength is increased, e.g., the ZFC and FC data split at 5~K in 5~T. Additional small peak is observed in the FC 0.1 T and 0.2 T data around 6 K, which disappears in stronger fields.

At room temperature, a single paramagnetic ESR line was observed. The $g$ factor and its anisotropy were obtained by recording the line position as the applied field orientation was varied with respective to the crystallographic axes. The upper-left inset of Fig.~\ref{Fig3}(a) shows the obtained $g$ factor anisotropy, which exhibits a characteristic $\cos^2\theta$ angular dependence as the field orientation varies from $H\parallel c$ to $H\perp c$. The principal values of the $g$ factor are found as $g({\parallel}c)$ = 2.37 and $g({\perp}c)$ = 2.24. The $g$ factor anisotropy in the $ab$ plane was smaller than the ESR linewidth of 20~mT which gives an upper bound of 0.05 for the $ab$ plane anisotropy.

\begin{table}[b]
	\caption{Valence states in octahedral and triangular pyramid environment for the different Ir ions.}
	\begin{tabular}{c|c|c}
		\hline\hline
		& Ir1   ~~    Ir2   ~~    Ir3 & Ir4\\
		\hline
		~~A~~	&	~~3+   ~~  	5+   ~~  	6+~~	&	~~5+~~\\
		~~B~~	&	~~5+   ~~ 	4+   ~~  	5+~~ 	&	~~5+~~\\
		~~C~~	&	~~4+   ~~   6+	 ~~     4+~~ 	&	~~5+~~\\
		\hline\hline
	\end{tabular}\label{table2}
\end{table}

It is surprising that this compound displays an average effective moment of 2.1 $\mu_\mathrm{B}$, when similar compounds have been reported to exhibit effective moments from 0.2 to 1.3 $\mu_\mathrm{B}$ \cite{Nag15arxiv,Sakamoto06JSSC}. One possible scenario would be that all 4 iridium sites exhibit unusually high moments of $\sim$2 $\mu_\mathrm{B}$. Below we present another possible interpretation, which merits further experimental and theoretical consideration. Let us start from considering possible valence states for the different iridium environments in $\mathrm{Ba_{21}Ir_9O_{43}}$. The most common valence states for Ir ions in oxides are Ir$^{3+}$, Ir$^{4+}$, Ir$^{5+}$, and Ir$^{6+}$, which may apply to Ir1, Ir2, and Ir3 sites. For the Ir4 sites with the dimers of face-sharing octahedra, one can make reference to the data from a large family of $\mathrm{Ba_3}M\mathrm{Ir_2O_9}$, where $M$=Mg, Ca, Sc, Ti, Zn, Sr, Zr, Cd and In \cite{Sakamoto06JSSC, Nag15arxiv}, which have structurally very similar dimer units. In this family, a systematic relation is found between the valence state and Ir-Ir as well as Ir-O distances of the face-sharing octahedra, i.e., larger Ir-Ir and smaller Ir-O distances for higher valence states \cite{Sakamoto06JSSC}. For instance, the Ir$^{5+}$ states are stabilized in $\mathrm{Ba_3ZnIr_2O_9}$ with the Ir-Ir distance of 2.75 {\AA}, and similarly for others with $M$=Mg, Cd, and Sr \cite{Sakamoto06JSSC, Nag15arxiv}. The very similar Ir-Ir distance of 2.74 {\AA} in $\mathrm{Ba_{21}Ir_9O_{43}}$ indicate that Ir$^{5+}$ states are realized on Ir4 sites.

Next we can list possible valence states for Ir ions by imposing the charge neutrality condition for the overall system. For the given chemical formula $\mathrm{Ba_{21}Ir_9O_{43}}$, we find that only three valence configurations are possible \cite{SM}, which are listed in Table~\ref{table2}. For the two regular octahedral sites, Ir1 and Ir3, it is found that their Ir-O distances are very similar, 1.95(3) {\AA} and 1.90(3) {\AA}, respectively. This indicates that they likely adopt the same valence state, excluding the combination A. To discriminate between combinations B and C, we consider the charge balance within the layer II. As can be seen from Fig.~\ref{Fig1}, there are no shared oxygens between neighbouring iridium ions within a layer, with Ba$^{2+}$ ions being found between Ir-O clusters. For B the charge on octahedral units (IrO$_6$ clusters) is $Q_\mathrm{oct}^\mathrm{B} = -7$, while on a triangular pyramid (IrO$_4$ cluster) it is $Q_\mathrm{pyr}^\mathrm{B} = -4$ \cite{calculation}. However, for C these values become much more disproportionate, $Q_\mathrm{oct}^\mathrm{C} = -8$ and $Q_\mathrm{pyr}^\mathrm{C} = -2$, which appears rather unlikely. Therefore, we conclude that the combination B is the most likely valence configurations.

We can now turn our attention to the consideration of magnetic moments residing on individual iridium sites. The values of effective moments for the case B, for instance, are calculated by using
\begin{equation}
(\mu_\mathrm{eff}/\mu_\mathrm{B})^2N_\mathrm{Ir} = \mu_\mathrm{pyr}^2N_\mathrm{pyr}+\mu_\mathrm{oct}^2N_\mathrm{oct}+\mu_\mathrm{dim}^2N_\mathrm{dim},
\end{equation}
where $N_\mathrm{Ir}$ is the total number of Ir sites, $N_\mathrm{pry}=1$, $N_\mathrm{oct}=2$, and $N_\mathrm{dim}=6$ are respectively the numbers of sites for $\mathrm{Ir^{4+}}$ in triangular pyramid, $\mathrm{Ir^{5+}}$ in octahedra, and $\mathrm{Ir^{5+}}$ in dimer. If as predicted octahedrally coordinated Ir$^{5+}$ is in a spin-orbit coupled non-magnetic state \cite{Khomskii}, all the magnetic moment deduced from the Curie constant must be assigned to the Ir$^{4+}$ ions on Ir2 site. This would lead to a rather large effective moment of 6.15 $\mu_\mathrm{B}$. This value is only slightly reduced by adopting the more likely literature values for Ir$^{5+}$ discussed above. For instance, when 0.2 $\mu_\mathrm{B}/\mathrm{Ir}$ for the dimers \cite{Nag15arxiv} and 0.91 $\mu_\mathrm{B}/\mathrm{Ir}$ for the regular octahedra \cite{Cao14PRL} are used, the moment expected for each Ir$^{4+}$ becomes 6.0 $\mu_\mathrm{B}$ which is still fairly large. Interestingly it is close to the expected value of 5.9 $\mu_\mathrm{B}$ for a state $S=5/2$.

To the best of our knowledge, Ir$^{4+}$ within the triangular pyramid environment has not yet been reported or considered theoretically. Our preliminary {\it ab initio} many-body configuration interaction calculations \cite{Katukuri12PRB} on IrO$_4$ unit, performed to understand the electronic multiplet structure of the Ir$^{4+}$ ion at Ir2 site, indicate an orbitally nondegenerate low-spin ground state. As a result, the spin-orbit coupling in the 5$d$ Ir atom, though strong, has little effect on the ground state - less than 20\% contribution from the excited orbital states. However, a distortion, e.g., the Ir$^{4+}$ atom moving towards the tetrahedral position of the triangular pyramid, could help result in a high-spin $S=5/2$ ground state. In fact, our calculations on an ideal IrO$_4$ tetrahedra with Ir-O bond lengths of 1.9 \AA ($\approx$ average of the Ir-O bond lengths at the Ir2 site) show that the $^6A_1$ ($S=5/2$) state is rather close, $\sim 0.33$ eV, to the low-spin $^2T_2$ ground state of the tetrahedral point group symmetry. The high-spin state could then be stabilized by further distortion, interactions, or other correlated effects, which we leave for future work.

\begin{acknowledgments}
We thank P. Huang and P. Babkevich for helpful discussions. This work was supported by the Swiss National Science Foundation, its Sinergia network MPBH, and European Research Council grant CONQUEST. M.J. is grateful to support by European Commission through Marie Sk{\l}odowska-Curie Action COFUND (EPFL Fellows).
\end{acknowledgments}


\begin{thebibliography}{99}
\bibitem{calculation} For both the Ir1 and Ir3 octahedral units in combination B, each Ir$^{5+}$ ions are surrounded by six O$^{2-}$ ions so that the total charge ($Q_\mathrm{oct}^\mathrm{B}$) sums up as $5+6\times(-2)=-7$. Likewise, for the Ir2 triangular pyramidal unit, each Ir$^{4+}$ ions surrounded by four O$^{2-}$ ions provide the total charge ($Q_\mathrm{pyr}^\mathrm{B}$)  of $4+4\times(-2)=-4.$
\end{thebibliography}
\end{document}